\documentclass[%
 reprint,
 amsmath,amssymb,
 aps,
]{revtex4-2}
\usepackage{subfigure}
\usepackage{overpic}
\usepackage{graphicx}
\usepackage{dcolumn}
\usepackage{bm}

\begin{document}
\bibliographystyle{unsrt}
\preprint{APS/123-QED}

\title{A Double-Spring Model for Nanoparticle Diffusion in a Polymer Network }

\author{Yu Lu}
\affiliation{School of Mechanics and Engineering Science, Shanghai Institute of Applied Mathematics and Mechanics, Shanghai Key Laboratory of Mechanics in Energy Engineering, Shanghai University, Shanghai, China}
\author{Xin-Yue Liu}
\affiliation{School of Mechanics and Engineering Science, Shanghai Institute of Applied Mathematics and Mechanics, Shanghai Key Laboratory of Mechanics in Energy Engineering, Shanghai University, Shanghai, China}
\author{Guo-Hui Hu}
\email{ghhu@staff.shu.edu.cn}
\affiliation{School of Mechanics and Engineering Science, Shanghai Institute of Applied Mathematics and Mechanics, Shanghai Key Laboratory of Mechanics in Energy Engineering, Shanghai University, Shanghai, China}

\date{\today}

\begin{abstract}
The transport of nanoparticles (NPs) in polymer networks, as a typical simplified model describing various structures in living systems, is profoundly important in biomedical engineering and nanotechnology. Predicting the effective diffusivity of NP confined in an ordered network has been an intriguing focus in this frontier field.
In the present study, the diffusion of NPs in an unentangled polymer network for different NP radii and network stiffness is numerically investigated by single particle dissipative particle dynamics (DPD).
It is found that, the deformation due to the junction deviation contributes significantly to the the potential barrier $U$ for the NP to overcome during hopping, and it is dominated over the strain energy induced by loop stretching for larger NPs and lower network rigidity.
Analyses based on the theory of continuum mechanics reveal that the relation between this deformed energy and the junction deviation can be described by a non-linear spring.
Taking into account both effects of the loop stretching and junction deviation, a double-spring model is proposed to characterize the diffusivity of the NPs in the ordered network.
The theoretical prediction is in good agreement with our numerical simulations, and qualitatively consistent with the investigations available.
This model is helpful to improve our understanding on the dynamic behavior of nanoparticle in complex biological environment, and provide theoretical guidance in designing biomedical applications.

\end{abstract}

\maketitle

The diffusion of nanoparticles (NP) in heterogeneous medium, such as polymer liquids \cite{xue2016probing,kalathi2014nanoparticle,Ge2018Nanorheology,amblard1996subdiffusion,van2004Brownian,dong2015diffusion,yamamoto2015microscopic,Volgin2017Molecular,Dell2013Theory,Xue2020Diffusion}, porous medium \cite{godec2014collective,rahalkar2017diffusion,senanayake2019diffusion,Cherstvy2019Non,Banks2016Characterizing,Kumar2019Transport,toyota2011non,Johann2018Particle,Lu2021potential} is of great significance in targeting delivery of nanomedicines, various fundamental biological processes and plenty of physical chemical applications \cite{Turci2021Wetting,Garamella2020Anomalous,wang2013bursts,hofling2013anomalous,he2016dynamic,wang2009anomalous,wang2012brownian,Pastore2021Rapid}.
These heterogeneous media are usually described as local confinements in theoretical models\cite{Xue2020Diffusion,Dell2013Theory,Lu2021potential,cai2015hopping}, leading to the diffusion properties distinct from the classic Brownian motion.
Previous investigations \cite{Lu2021potential,Chen2020Nanoparticle} indicated that at short timescales the particles show a transition from ballistic to Brownian motion, while at intermediate time the NPs’ motion is restricted by the surrounding structures, manifesting a sublinear mean square displacement (MSD). At long timescales, the NP in complicated medium behaviors like Brownian motion with an effective diffusivity $D^L$.

\begin{figure*}[t]
	\centering
	\subfigure{
	\includegraphics[height=5.0cm]{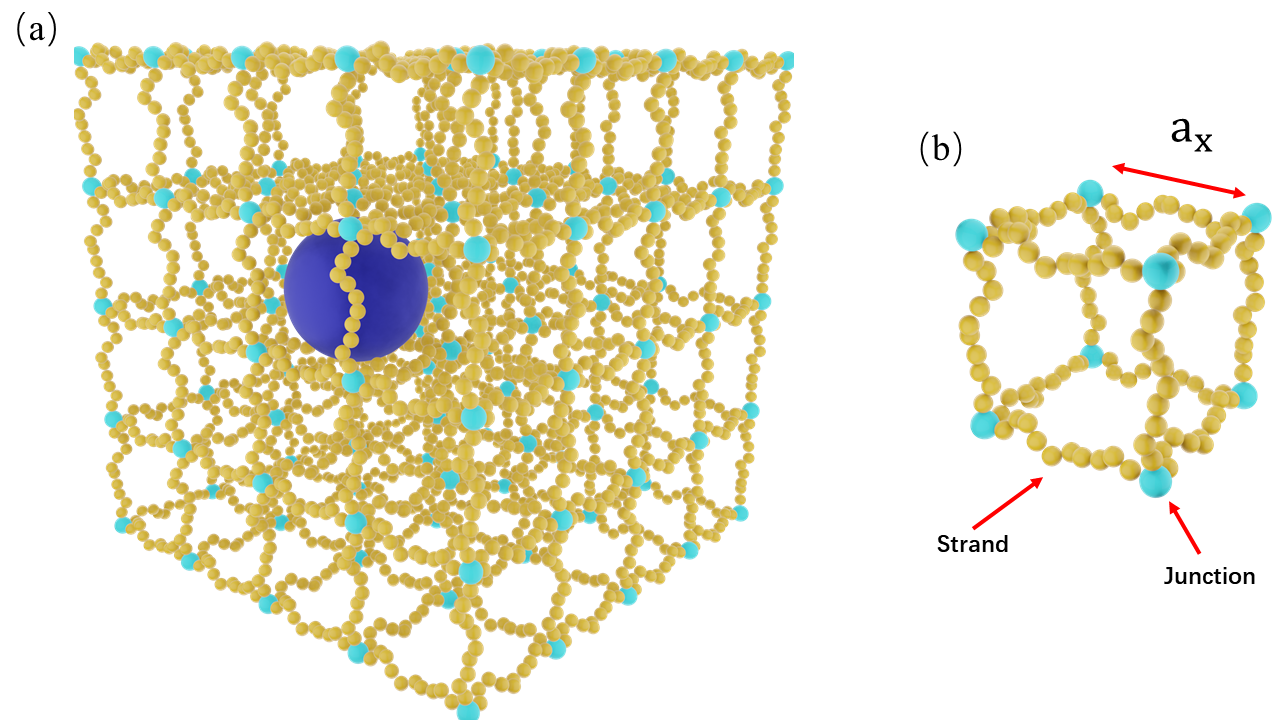}
	}
	\subfigure{
	\includegraphics[height=6.0cm]{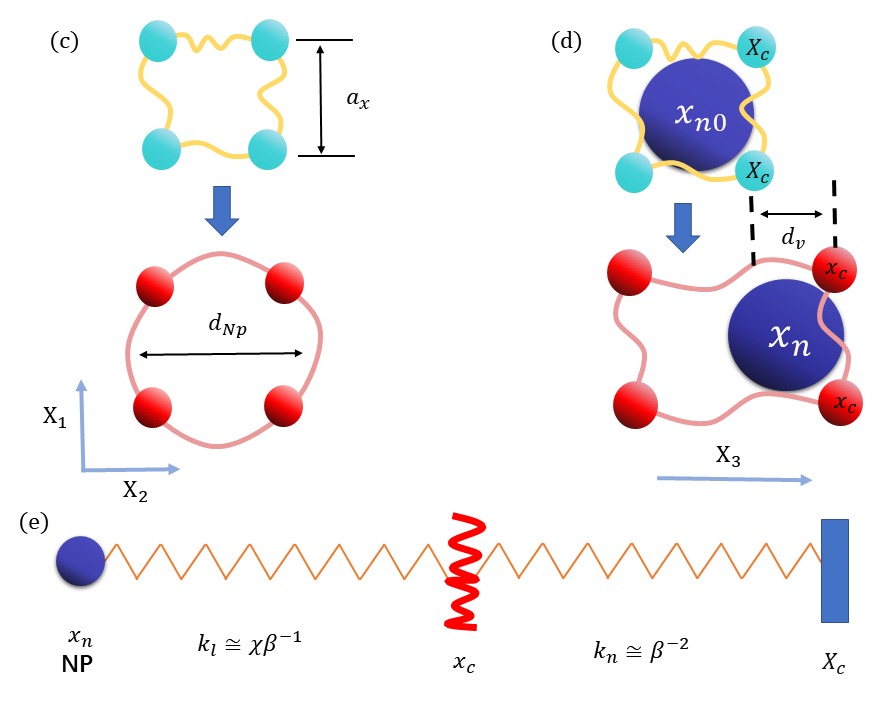}
	}
	\caption{
	(a) A schematic depiction of the ordered network constructed by polymer strands and NP (blue).
	(b) The average distance between two junctions (cyan), which is connected by strand (yellow), is $a_x$.
	A double spring model (e) is proposed to simultaneously characterize the deformation of the polymer network and the potential barriers overcome by the nanoparticle during hopping.
	The first spring measures the loop stretching (c) in the mesh of NP occupied, which can be simplified by a linear spring with stiffness $k_l \simeq \chi \beta^{-1}$.
	The junction deviation (d) when NP tries to squeeze in the polymer loop will induce another kind of deformation in polymer network.
	By using the continuum mechanics, the deformation energy of junction deviation can be represented by a sublinear spring with stiffness $k_n \simeq \beta^{-2}$.
	}
	\label{img:double_spring}
\end{figure*}

The motion of a NP in a heterogeneous medium with strong local confinement, such as permanently cross-linked polymer network, has been described by the nonlinear Langevin equation (NLE), and the local constraints can be considered as a potential barrier $U$ \cite{Dell2013Theory,Xue2020Diffusion,cai2015hopping,Lu2021potential}.
Activated hopping of nanoparticles
can be observed \cite{wang2009anomalous,cai2015hopping,Dell2013Theory,xue2016probing,Xue2020Diffusion,Lu2021potential} when it diffuses by overcoming the free energy barrier.
The mechanism of NP hopping in polymer networks is worth investigating in depth, which might be closely related to the physical origin of the potential barrier. However, the theories available\cite{cai2015hopping,Dell2013Theory} presented different scaling law of the potential barrier dependent on the NP radius.
Cai {\it et al.}\cite{cai2015hopping} suggested the mechanism underlying is the loop stretching in polymer networks, and proposed a scaling laws based on the ``elementary'' network theory under the assumption that the restoring force of the loops is linearly elastic with the stiffness proportional to the NP size, and the deformed energy can be written in a parabolic form.
Dell and Schweizer\cite{Dell2013Theory} developed a theory of hopping based on a nonlinear Langevin equation and polymer reference interaction site model (PRISM)
\cite{Hooper2005Contact}, showing that the potential barrier is linearly increase with NP size, following by a cubical increase.
Utilizing molecular dynamics simulations, Sorichetti {\it et al.} \cite{Sorichetti2021Dynamics} reported that both theories present satisfying predictions, and it is hard to evaluate their reliability.
The numerical study conducted by Chen {\it et al.}\cite{Chen2020Nanoparticle} found that the diffusion behaviors of NP could be qualitatively different as the length of polymer chains in the network varies. Using the strand fluctuation distance $d_f$ as a normalized length scale, they found that the polymer networks with shorter polymer strands (lower $d_f$) perform higher local confinement on NPs, which corresponds to larger potential barrier.

Although extensive investigations have been conducted to explore the NP diffusion in polymer networks, a significant question remaining open is to predict the diffusive behavior theoretically. An effective way to solve this problem might be to formulate an accurate mathematical expression for the potential barrier in nonlinear Langevin equation, which requires an in-depth understanding of the deformation of the polymer network when the nanoparticle traverses its loop. As indicated by previous literatures \cite{cai2015hopping,Lu2021potential}, the potential barrier needed to be overcome by the NPs is related to the deformation energy of the polymer network as the NPs hopping. Thus two crucial questions have to be examined. The first one is to quantitatively understand the physics behind the polymer network deformation during NP squeezing the loop. This goal can be achieved by analyzing the numerical information based on mesoscopic simulations, whereas would be difficult for experimental measurements. Secondly, how the properties of the polymer network, especially mechanical properties, influence the local confinement on NP? The theories developed by James \& Guth\cite{James1947Statistical} and Flory\cite{Flory1977Theory} are helpful to connect these two questions. In their theory, the positions of junctions in polymer network is in the form of Gaussian probability density function, and their standard deviations are defined as the fluctuation length $d_f$.
Then the elasticity modulus $\nu$ of polymer network 
has magnitude of $\nu \sim d_f^{-2}$\cite{Rubinstein1997Nonaffine}.
In our recent numerical study\cite{Lu2021potential}, we carried out a detailed analysis on the trajectories of NP and the junctions in polymer network.
It is found that the junctions in polymer network fluctuate around their origin locations with a non-zero mean value when NP tries to hop in the polymer loop,
suggesting that other than the loop stretching, the deformation energy due to the junction deviation during hopping contributes significantly to the potential barrier.

The single-particle Dissipative Particle Dynamics (DPD)\cite{hoogerbrugge1992simulating,Pan2008Single,pan2009rheology} is used to simulate the diffusion of NP in an unentangled polymer network in the present study.
As shown in Fig. \ref{img:double_spring}(a)(b), the network with cubic topology is constructed by the polymer strands modelled by the classical bead-spring model. 
A single strand in the polymer network is consisted of $N$ DPD particles, which represent the Kuhn monomer with size $b$. The polymer strands in the network are permanently connected by the junctions.
The average distance between two neighbouring junctions is the mesh size $a_x$, which is given by $a_x \simeq bN^{\mu}$, in which $\mu$ is the Flory exponent depending on the solvent properties. Usually $\mu = 0.588$ for athermal and good solvent.
The system is filled by the solvent molecules represented by single DPD particles, and their number density $n$ is kept to be a constant. The volume fraction of polymer network can be estimated by $\varphi=(3N-5)/na_x^3$. 
The correlation length $\xi$ can be calculated by $\xi \simeq b\varphi^{-\mu/(3\mu-1)}$\cite{cai2011mobility},
which yields the tube diameter of $a_e \approx 5\xi$\cite{cai2011mobility}.
In present work, $N$ and $\varphi$ are selected to keep the mesh size $a_x$ smaller than tube diameter $a_e$, thus the polymer networks are unentangled. Full details of the numerical techniques are described in the Supplement Information S1. 

Since the potential barrier in nanoparticle diffusion is closely correlated with the polymer network deformation, a normalized fluctuation length $\beta=d_f/a_x$ 
is introduced aiming at considering the influences of the stiffness on the deformation. 
It should be mentioned that there is a slight difference in definition of fluctuation length in previous literature. Flory {\it et al.}\cite{Flory1977Theory} proposed that the local constraints on junctions limit the fluctuation of strands, which affects the elasticity of polymer network. They considered $d_f$ as the mean square displacement(MSD) of the junctions, while Chen {\it et al.}\cite{Chen2020Nanoparticle} defined it as the MSD of the central monomer of strands. It is shown in Supporting Information S2 that the results obtained from these two definitions are similar for the unentangled polymer network, and the formulation of Flory {\it et al.}\cite{Flory1977Theory} is used in the present study. 

\begin{figure}[]
	\centering
	\subfigure{
		\begin{overpic}[height=4cm]{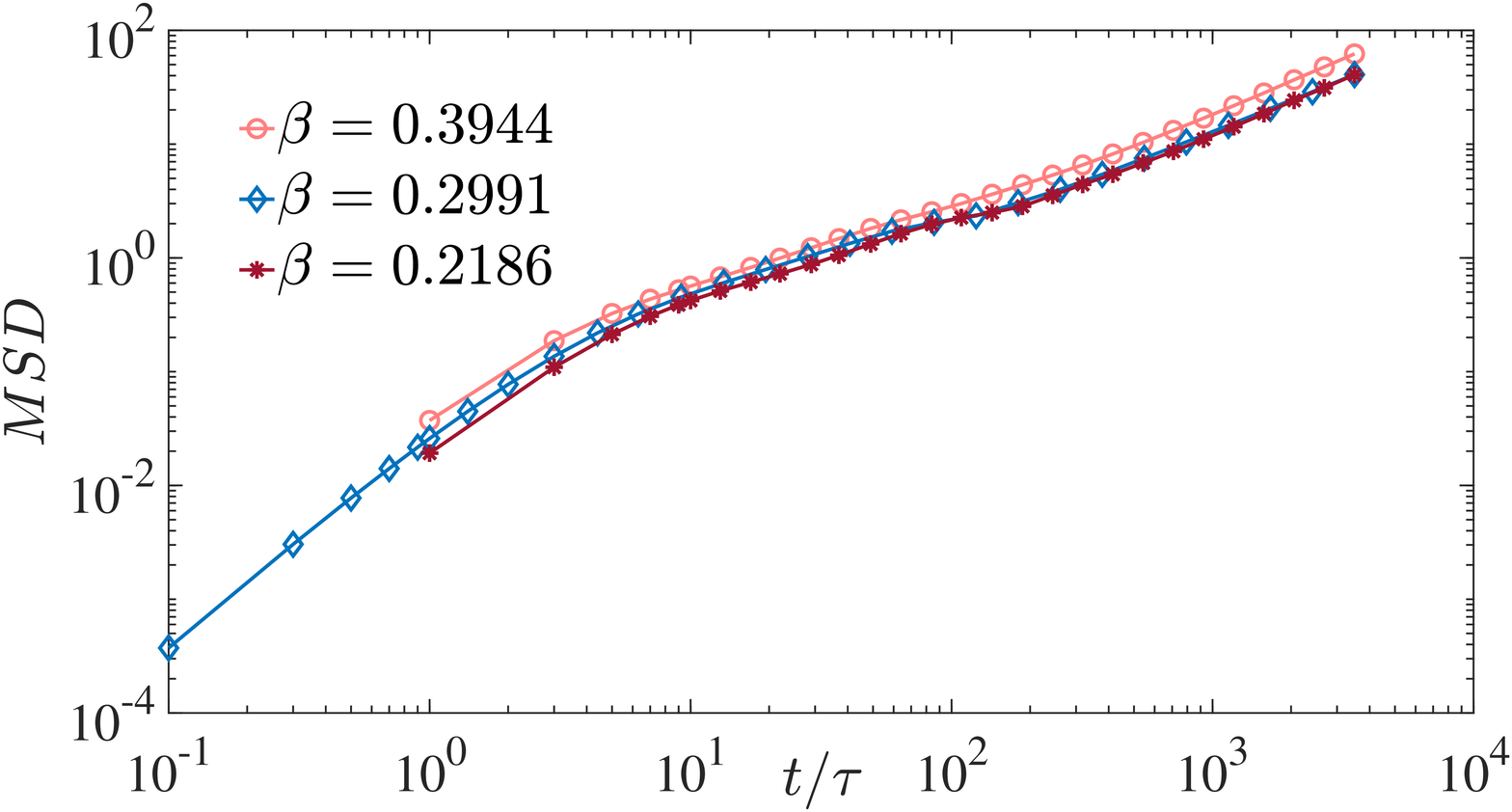}
            \put(84,42){(a)}
        \end{overpic}
        }
	\subfigure{
		\begin{overpic}[height=4cm]{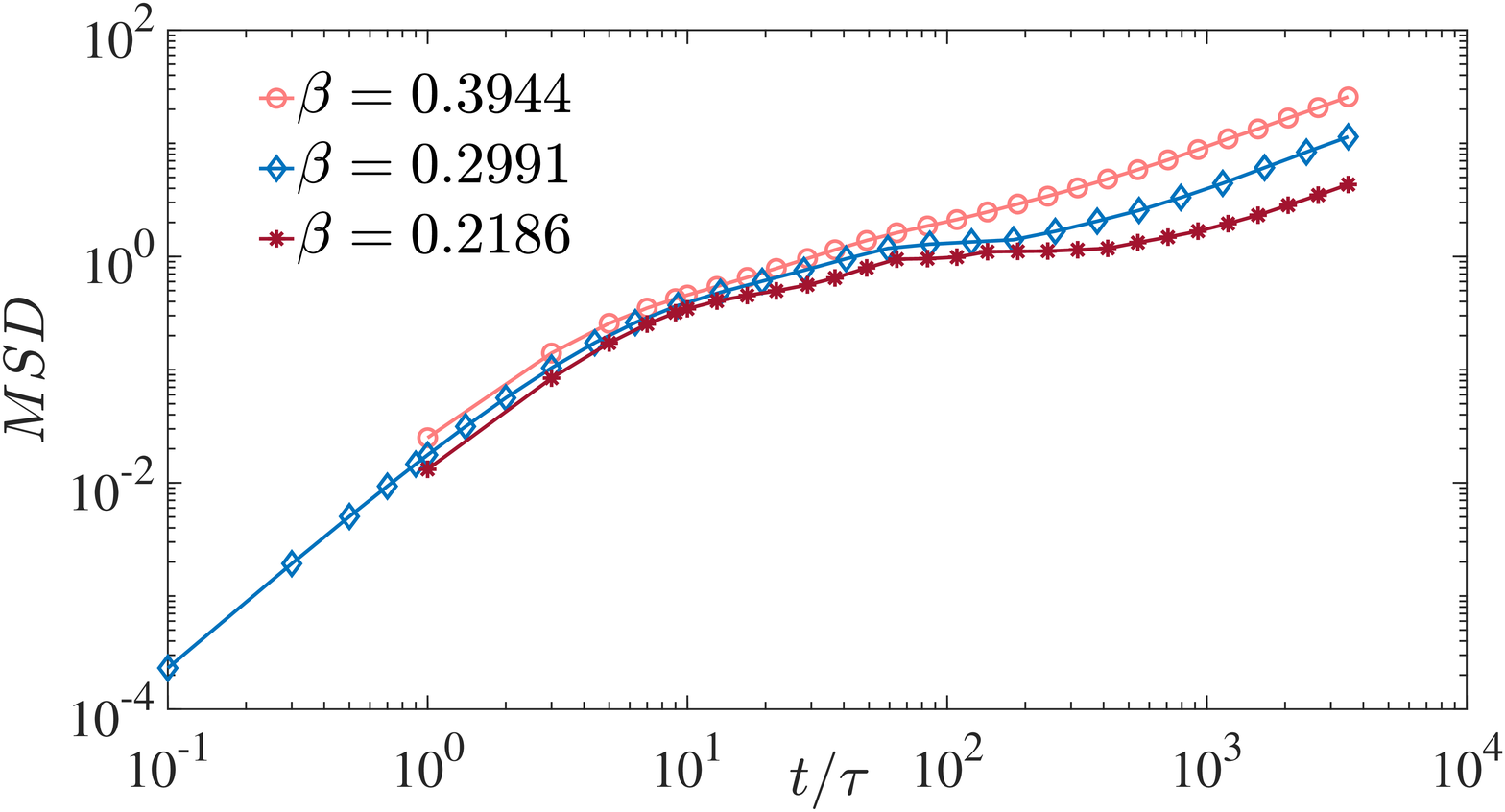}
            \put(85,42){(b)}
        \end{overpic}
	}
	\subfigure{
		\begin{overpic}[height=4cm]{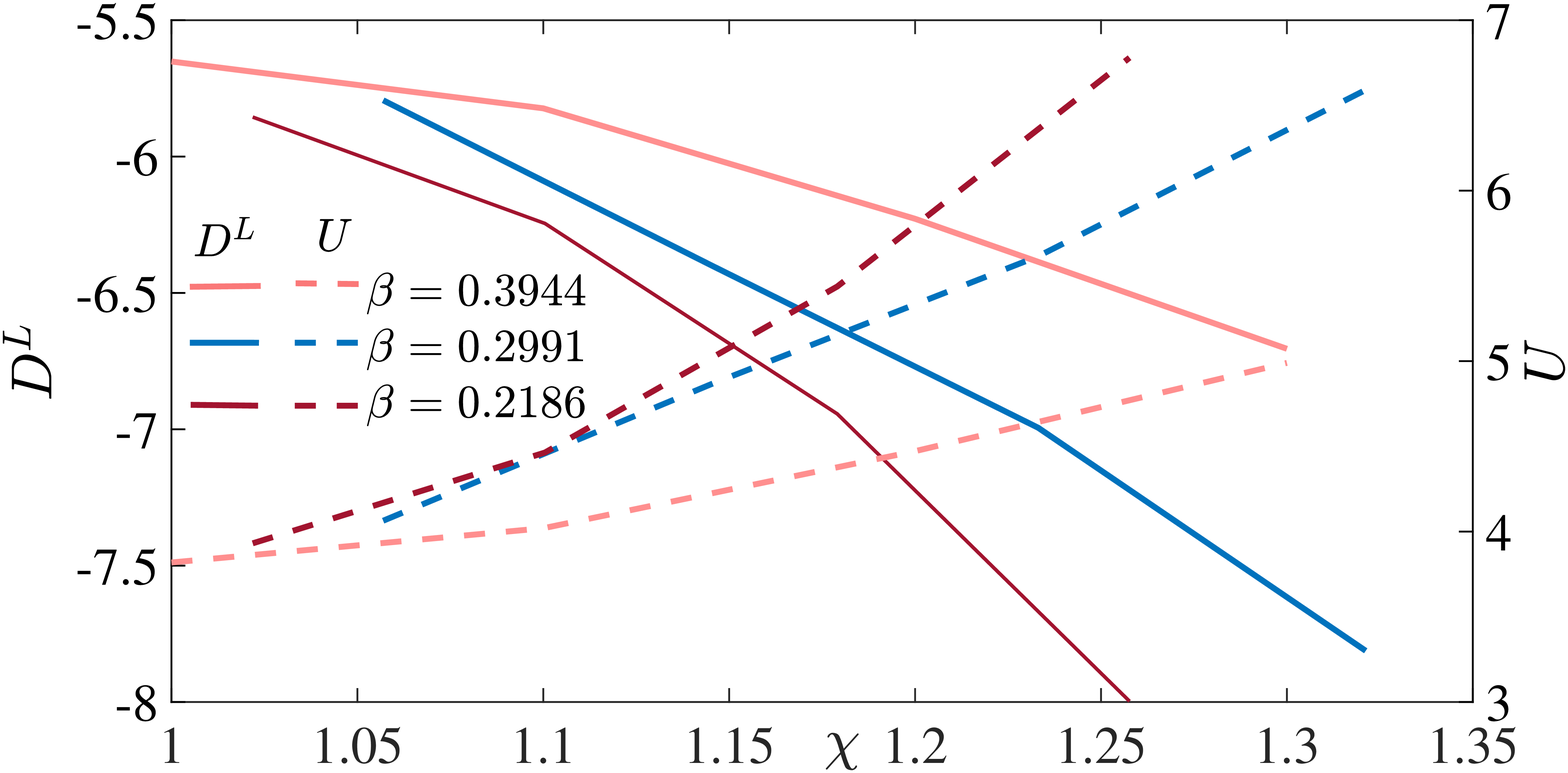}
            \put(85,40){(c)}
        \end{overpic}
	\label{img:U_and_Dr}
	}
	\caption{
        MSD for smaller NP size $\chi=1.1$(a) and larger NP size $\chi=1.3$(b) varied by normalized fluctuation length $\beta$ is plotted against time in log-log coordinate. (c)
        The potential barrier (right, dash line) for NP to overcome and the corresponding long time diffusivity (left, solid line) varied by normalized fluctuation length $\beta$ are depicted against the normalized NP size $\chi$.
	}
	\label{img:MSD_vs_axae}
\end{figure}

As shown in Fig. \ref{img:MSD_vs_axae}, the mean square displacement of NP varying with the normalized NP size $\chi=d_{NP}/a_x$ is depicted against time in log-log coordinates.
It is found that, at the short-time stage, the dynamics of NP is dominated by inertia, the ballistic motion can be observed by $MSD \simeq t^2$ when the NP has not yet collided with polymer strands.
At the intermediate stage, the local confinement constructed by polymer network limits the diffusion of NPs, and the MSD is sublinear with elapsed time.
In the long-time stage, the NP walks randomly cross the polymer network and MSD becomes linear, which gradually returns a long-time diffusivity $D^L$.

The NP diffusion in polymer network exhibits distinct dynamic behavior as the fluctuation length $\beta$ differs.
For smaller NP with $\chi=1.1$ (Fig .\ref{img:MSD_vs_axae}(a)), although the discrepancy of nanoparticle MSD between different $\beta$ is negligible for both intermediate- and long-time stage\cite{Chen2020Nanoparticle}, it is slightly greater for larger fluctuation length $\beta$.
For the larger NP with $\chi=1.3$ (Fig .\ref{img:MSD_vs_axae}(b)), smaller $\beta$, corresponding to larger stiffness of the polymer network according to Flory's theory\cite{Flory1977Theory}, leads to stronger local confinement on NP, which can be observed by a plateau in the MSD for intermediate-time stage. 
This implies the occurrence of hopping phenomenon, and results in the rapid reduction of the long time diffusivity $D^L$ if the nanoparticle radii $\chi$ increases, as shown in Fig. \ref{img:U_and_Dr}.
The intermediate- and long-time behaviors of NP are qualitatively different as $\beta$ varies.
Similar results have been reported by Chen {\it et al.} recently\cite{Chen2020Nanoparticle}.

Utilizing the Nonlinear Langevin equation (NLE) and Kramer’s theory, it will yield the effective diffusivity once the potential barrier $U$ is obtained. For strong local confinement ($U>3$), it has a simplified form given by $D^L \sim e^{-U}$\cite{Dell2013Theory,Xue2020Diffusion,Lu2021potential}. As demonstrated by previous study\cite{Lu2021potential}, the deformations due to both loop stretching and junction deviation contribute to the potential barrier $U$ for NP to overcome, as shown in Fig. \ref{img:double_spring}, and the prediction of the effective diffusivity based on the potential barrier $U$ calculated according to this idea is in good agreement with the simulation results of DPD. 
For the influences of the network stiffness on NP diffusion, it is shown in Fig. \ref{img:U_and_Dr} that the potential barrier is generally larger for smaller $\beta$, and consequently the effective diffusivity will be smaller. 
Furthermore, as the fluctuation length becoming smaller, $U$ grows more rapidly with $\chi$.
Therefore, both normalized NP radius $\chi$, which is widely used as the measurement of the confinement by previous study\cite{cai2015hopping,Dell2013Theory,senanayake2019diffusion}, and normalized fluctuation length $\beta$, determine the local confinement constructed by polymer network on NPs.

\begin{figure*}[t]
	\centering
	\subfigure{
		\begin{overpic}[height=5.0cm]{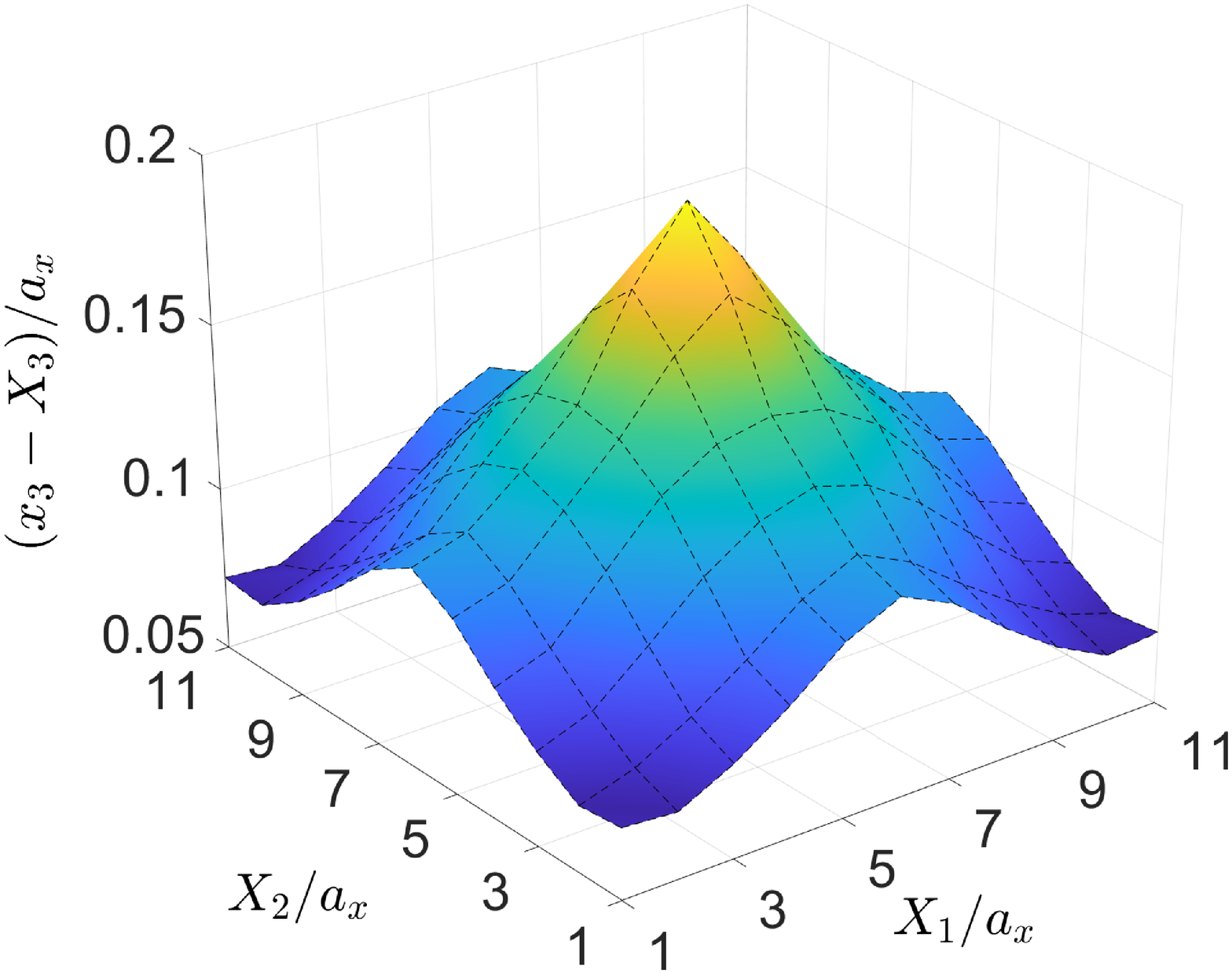}
            \put(24,50){(a)}
        \end{overpic}
	}
	\subfigure{
		\begin{overpic}[height=5.0cm]{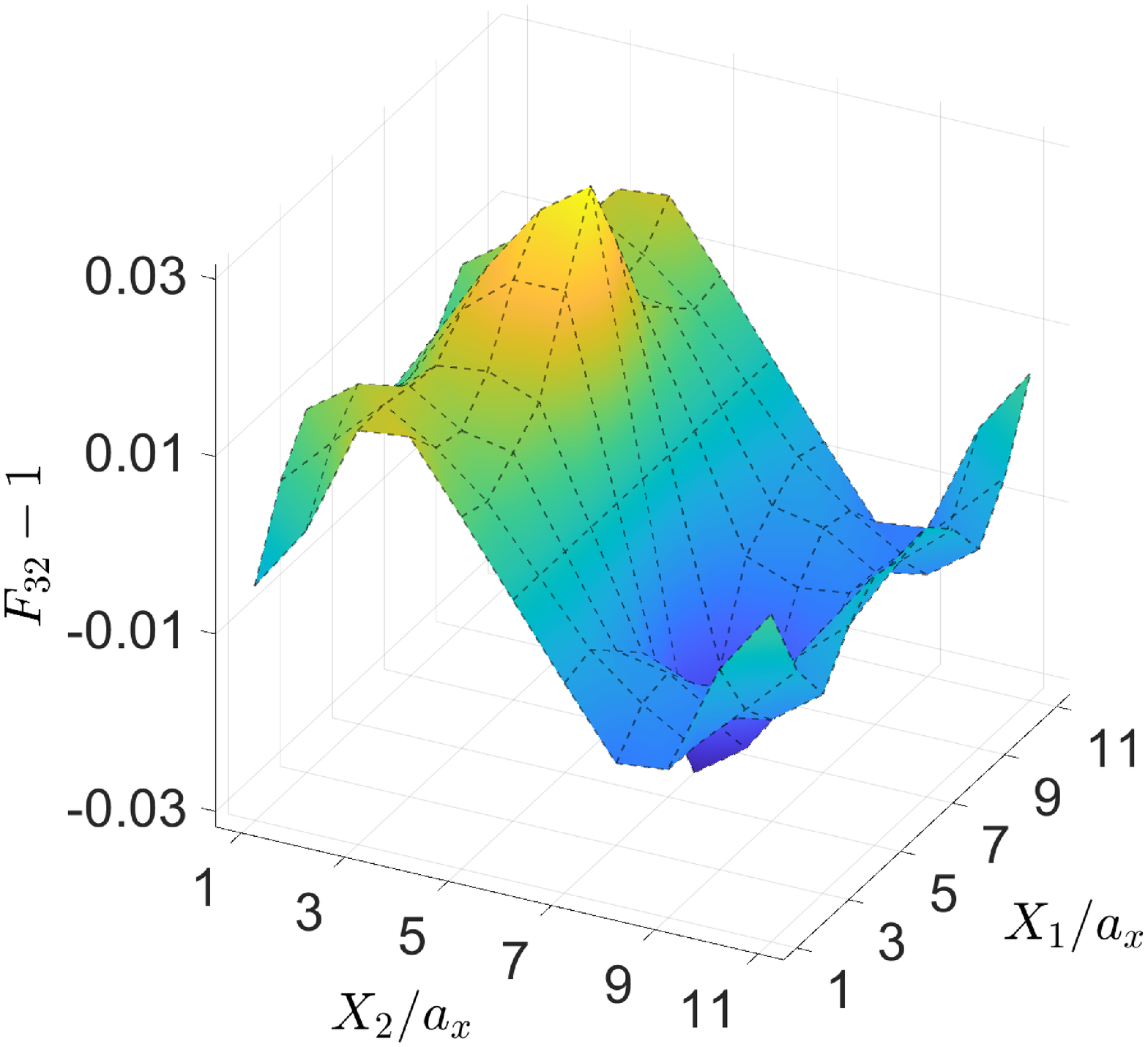}
            \put(26,50){(b)}
        \end{overpic}
	}
	\subfigure{
		\begin{overpic}[height=4cm]{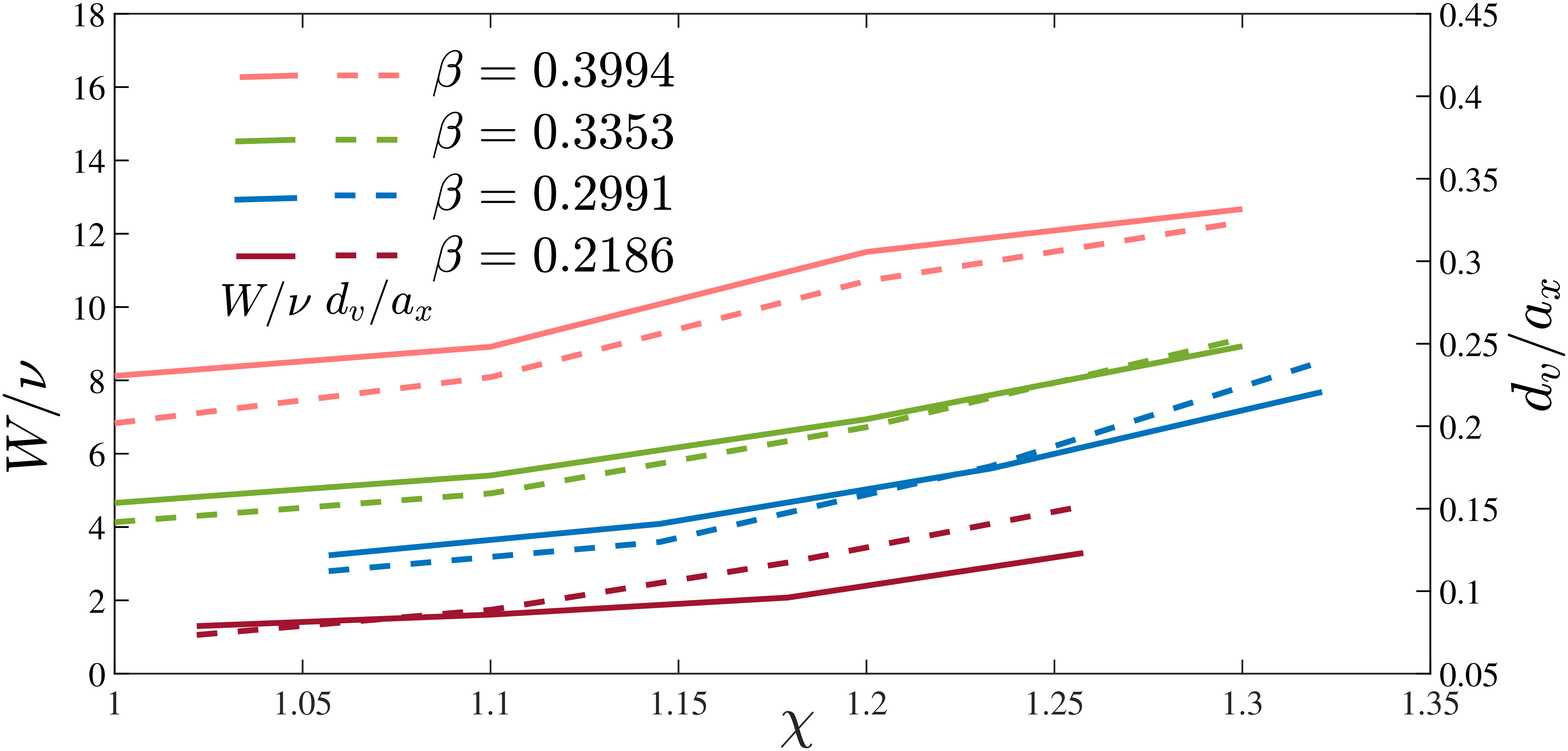}
            \put(80,40){(c)}
        \end{overpic}
	}
	\subfigure{
		\begin{overpic}[height=4cm]{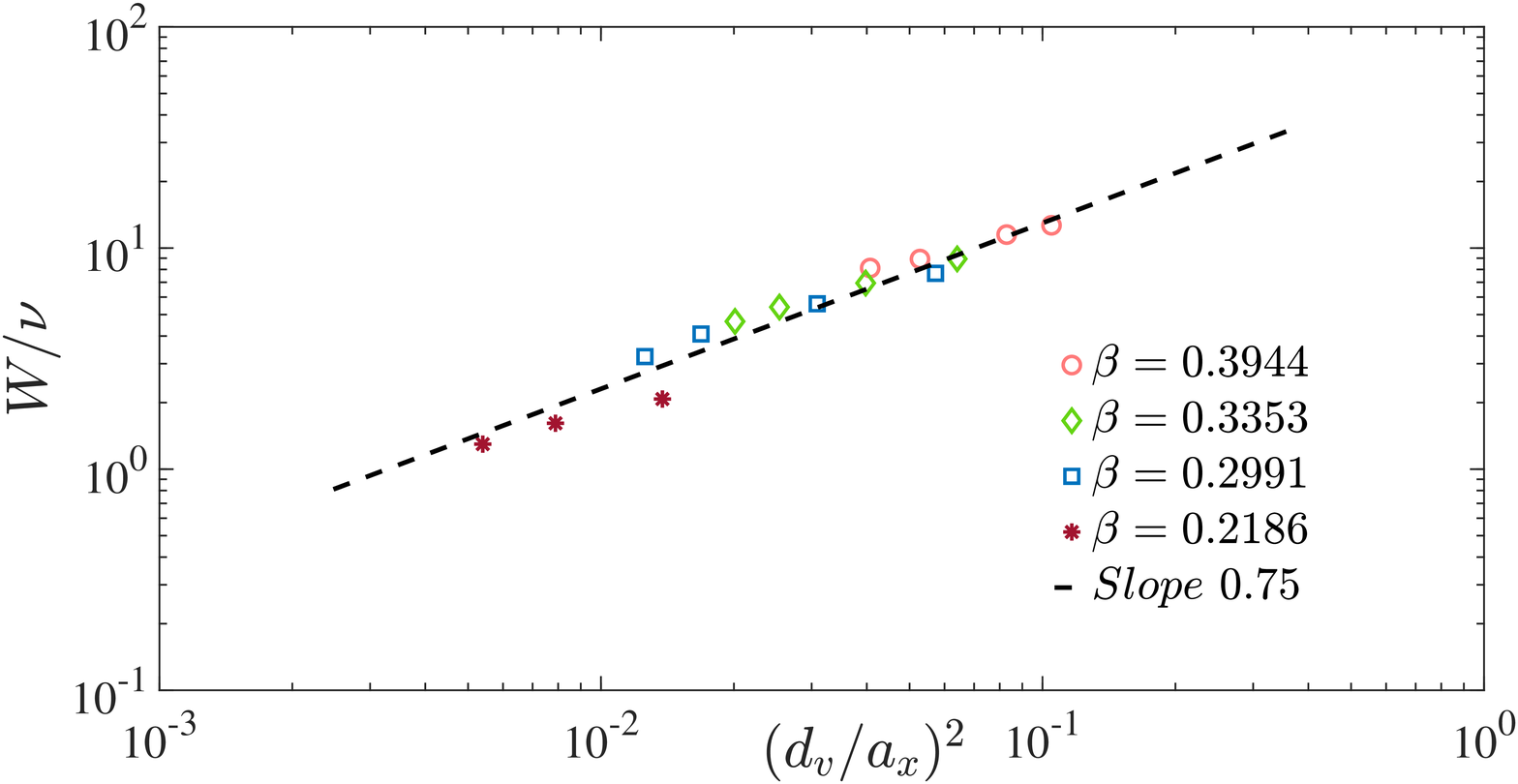}
            \put(14,40){(d)}
        \end{overpic}
	}
	\caption{(a) Displacement and (b) corresponding shearing strain of the slice in network during NP hopping. 
	(c) The variation of the the deformation energy $W/\nu$ (left, solid line) and the normalized deviation length $d_v/a_x$ (right, dash line) for different $\chi$ and $\beta$.
    (d) Deformation energy $W/\nu$ {\it vs.} the normalized deviation length $(\frac{d_v}{a_x})^2$ follows a power law, and the slope is $0.75$ in log-log coordinates.
    }
	\label{img:strain}
\end{figure*}

The potential barrier for NP to overcome during translocation between neighbouring meshes is physically originated from the deformation energy.
To measure the deformation in the framework of continuum mechanics\cite{James1947Statistical}, the coordinate {$\bf{X}$} of a junction in equilibrium locations is used as the reference states, which can be obtained by the ensemble average of the junction's location in the undeformed polymer network.
It fluctuates from the equilibrium location in Gaussian distribution with the variance $d_f^2$. 
The NP hopping will induce the deformation of the polymer network, meanwhile the junction {$\bf{X}$} moves to a place with coordinate $\bf{x(X)}$.
The deformation gradient is calculated by
\begin{equation}
    F_{iK}({\bf X})=\frac{\partial  x_i({\bf X}) }{\partial  X_K},
\end{equation}
in which subscripts $i$ and $K$ are $1,2,3$, representing for three dimensions.
Then the deformation energy normalized by $k_BT$ can be calculated by the well known phantom network model proposed by James and Guth\cite{James1947Statistical}, 
\begin{equation}\label{equ:W}
    W=\frac{\nu}{2} \int (\lambda_1^2+\lambda_2^2+\lambda_3^2-3)dV
\end{equation}
in which $dV$ is the volume differential. $\lambda_i$ is the eigenvalues of the deformation gradient tensor, which stands for the stretches of $dV$ in principal directions. The network stiffness is estimated by $\nu \sim \beta^{-2}$ for Gaussian chains\cite{Rubinstein1997Nonaffine,Gusev2019Numerical}. 

Focusing on the slice in the network where the NP passes, its deformation and the component $F_{32}$ of the deformation gradient are shown in the Fig. \ref{img:strain}(a)(b) for $\beta=0.2991$ and $\chi=1.1$.
Without loss of generality, we assume the hopping direction of NP is in $X_3$ direction and the NP locates at the center of the network.
Then the junction in the slice is deformed in the hopping direction with a normalized displacement $(x_3-X_3)/a_x$.
It is found that the displacement and its gradient field are in axis-symmetry.
The maximum displacement is denoted as $d_v/a_x$ for the center junctions ${\bf{X}}_c$, which is the deviation length found in our previous study\cite{Lu2021potential}.
An important fact is, in ($X_1$ - $X_2$) plane perpendicular to the nanoparticle traversal, the displacement is almost zero even inside the element occupied by NP. This suggests that the deformed energy due to the junction deviation is independent of the mechanism of the loop stretching proposed by Cai et al. \cite{cai2015hopping}
It is also found that, the components $F_{23}$ and $F_{32}$ of deformation gradient tensor $F_{iK}$ are the main contributions in this deformation energy.

As shown in Fig. \ref{img:strain}(c), the variation of both the normalized deviation length $d_v/a_x$ and the deformation energy shows similar tendency as $\chi$ and $\beta$ varying, i.e., they increase for larger nanoparticle and softer network, or larger $\chi$ and $\beta$.
Moreover, the deformation energy $W$ can be related to the deviation length by a power law with an exponent $\alpha$ independent of the fluctuation length $\beta$, $W/\nu \sim ({d_v}/{a_x})^{2\alpha}$, as shown in Fig. \ref{img:strain}(d).
Therefore, the relation between $W$ and the elongation $d_v/a_x$ can be simplified as a non-linear spring:
\begin{equation}\label{equ:WS}
    W(x_c)=\frac{k_n}{2} (\frac{d_v}{a_x})^{2\alpha},
\end{equation}
in which the stiffness of the spring $k_n$ is proportional to $\nu \sim \beta^{-2}$, and the nonlinear parameter $\alpha \approx 0.75$ is obtained by fitting.

\begin{figure}[]
	\centering
	\subfigure{
		\begin{overpic}[height=4cm]{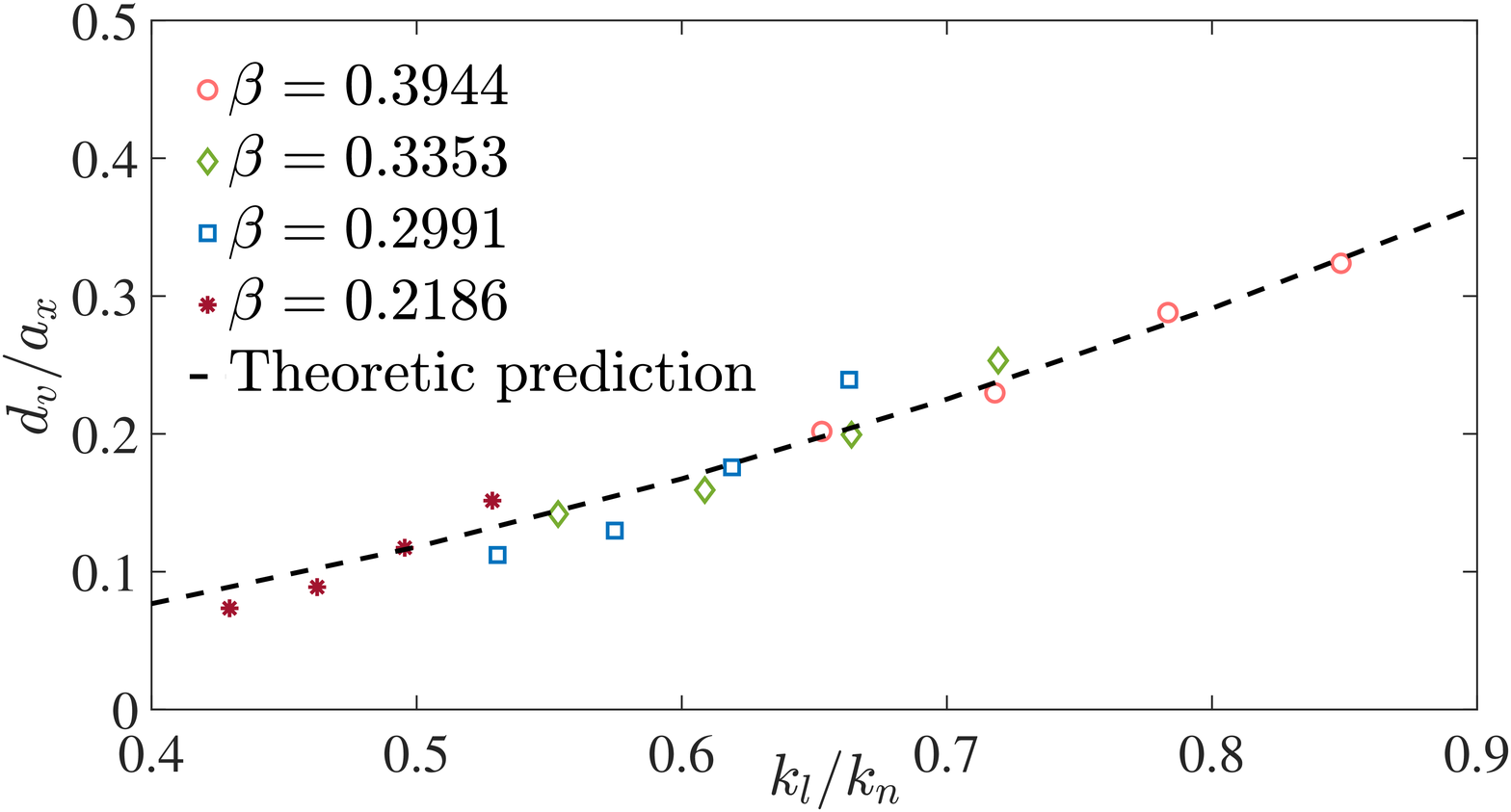}
            \put(80,42){(a)}
        \end{overpic}
	}
	\subfigure{
		\begin{overpic}[height=4cm]{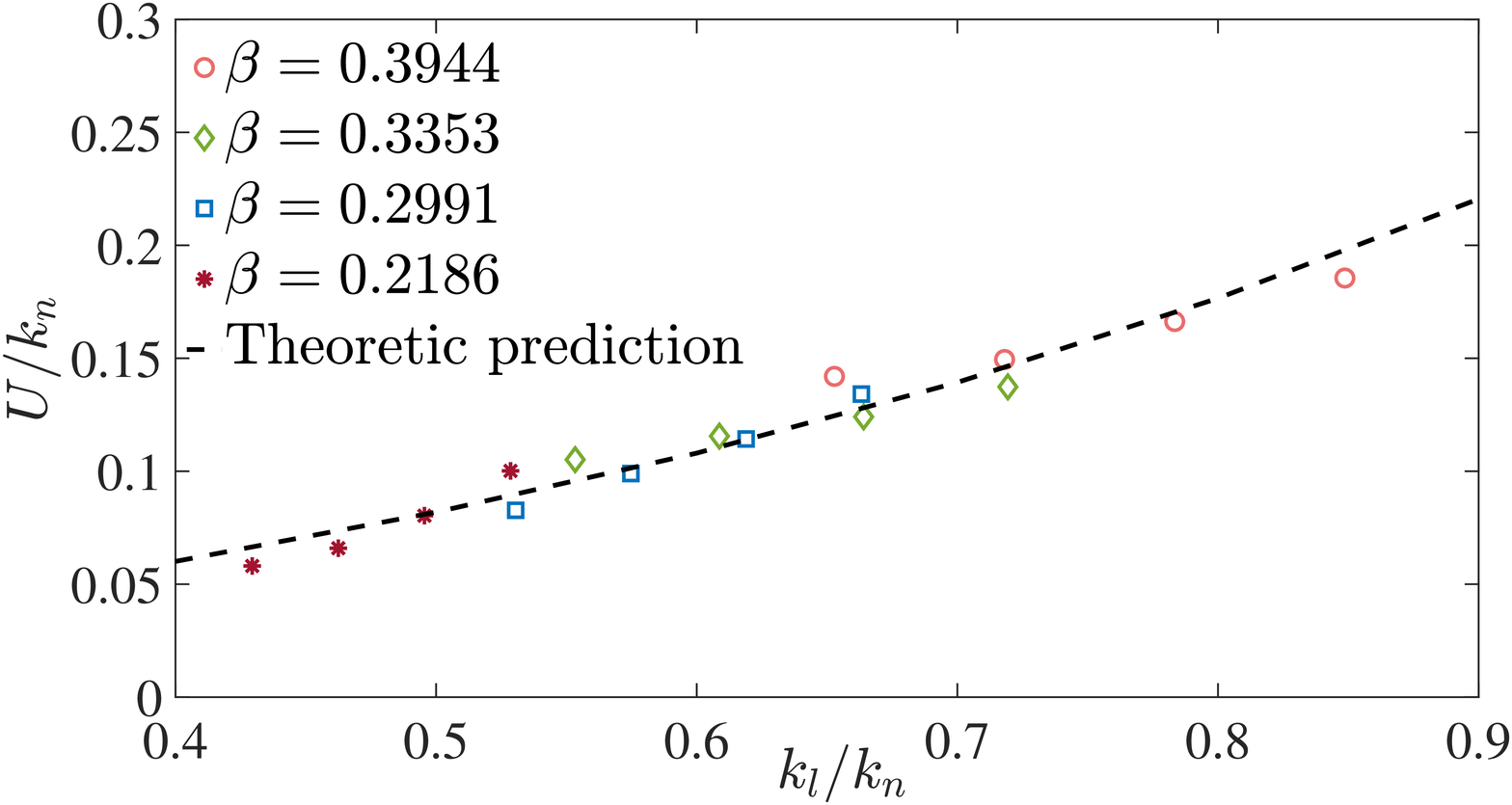}
            \put(80,42){(b)}
        \end{overpic}
	}
	\caption{The variations of the deviation length $d_v/a_x$ (a) and the potential barrier $U/k_n$ (b) with the ratio between the stiffness of two springs, which ensures the theoretical model (dash lines) will obtain diffusivity quantitatively agreed with the numerical simulations.
	}
	\label{img:Result_of_double_spring_model}
\end{figure}
These results indicate, there are two kinds of deformations contributing to the potential barrier during the NP diffusion, which are the loop stretching proposed by Cai et. al. \cite{cai2015hopping}, and the junction deviation due to NP squeezing, as shown in the Fig. \ref{img:double_spring}(d). In the first kind of deformation, the polymer loop in the mesh the NP occupies elongates in the $(X_1-X_2)$ plane, and its length scale changes from $a_x$ to $d_{NP}$. The second kind of deformation describes that, the particle locating at $x_{n0}$ moves to $x_n$ in $X_3$ direction, resulting in a displacement of the polymer strand by $d_v$, simultaneously the junctions originally locating at $X_c$ move to $x_c=X_c+d_v$.
Therefore, the potential barrier during hopping process can be characterized by a double-springs system (Fig. \ref{img:double_spring}(e)), which is given by
\begin{equation}\label{equ:TotalU}
    U(x_n)=U_l+W.
    \end{equation}
The first term $U_l$ is the deformation energy of loop stretching normalized by $k_BT$, which has been modelled as a linear spring, and can be written as (see Appendix C in Ref. \cite{cai2015hopping}),
\begin{equation}\label{equ:firstspring}
    U_l(x_n,x_c)=\frac{k_l}{2} (\frac{d_{nc}-d_{nc0}}{a_x})^{2},
\end{equation}
in which $d_{nc}=x_n-x_c$ is the elongation of this spring is the distance between NP and junctions. The equilibrium distance $d_{nc0}=x_{n0}-X_c$ is the half of mesh size $d_{nc0}=0.5a_x$. $k_l$ is the stiffness of the linear spring, which is considered to be proportional to the confinement degree of polymer network on the nanoparticle. 
The ratio $\chi$ between the NP diameter and the mesh size is a usual choice for confinement parameter. Alternatively, Chen {\it et al.}\cite{Chen2020Nanoparticle} suggested that the actual local confinement that NPs can feel is correlated to the fluctuation length of polymer strand $d_f$, rather than mesh size $a_x$.
So we assume that $k_l \sim \frac{d_{NP}}{d_f} =\chi \beta^{-1}$. The detailed derivation of equation (\ref{equ:firstspring}) is provided in the Supplement Information (S3).
The second term $W$ in equation (\ref{equ:TotalU}) is the strain energy induced by the junction deviation, which is described as the sublinear spring in equation (\ref{equ:WS}).
Substituting equation (\ref{equ:WS}) and (\ref{equ:firstspring}) into equation (\ref{equ:TotalU}) yields 
\begin{equation} \label{equ:instantaneous_U}
    \begin{split}
        U(x_n,x_c)&=U_l(x_n)+W(x_c)\\
        &=\frac{k_l}{2} (\frac{d_{nc}-d_{nc0}}{a_x})^{2} + \frac{k_n}{2} (\frac{x_c-X_c}{a_x})^{2\alpha}\\
    \end{split} 
\end{equation}
The equilibrium state of the system requires the free energy  $U(x_n,x_c)$ reaches a minimum, i.e.,
\begin{equation}\label{equ:equilibrium_assumption}
    \frac{\partial U(x_n,x_c)}{\partial x_c} =0.
\end{equation}
If we denote the arrival of the nanoparticle at the junction location $x_c$ as a successful hopping, i.e., $d_{nc}=x_n-x_c=0$, and consider the non-linear parameter $\alpha=0.75$, as shown in Fig. \ref{img:strain}(d), 
the  elastic coefficients of the two springs can be related by the deviation length $d_v/a_x$ using equation (\ref{equ:equilibrium_assumption}),  
\begin{equation}\label{equ:dv_dp}
    d_v/a_x=\frac{4}{9}(\frac{k_l}{k_n})^2
\end{equation}
Then the normalized potential barrier $U/k_n$ is
\begin{equation}\label{equ:U_dp}
    \frac{U}{k_n} = (U_l+W)/k_n = \frac{1}{8} \frac{k_l}{k_n} 
    + \frac{2}{9}(\frac{k_l}{k_n})^3 
\end{equation}
in which the parameters $k_l =A \beta^{-1} \chi$ and $k_n =B \beta^{-2}$ are determined by the properties of polymer network $\beta=d_f/a_x$ and normalized NP size $\chi=d_{NP}/a_x$\cite{Rubinstein1997Nonaffine,Chen2020Nanoparticle}.
Based on the potential barrier we obtained, the coefficients $A$ and $B$ can be estimated by utilizing a gradient descent algorithm and the least square method, as presented in Fig. \ref{img:Result_of_double_spring_model}(b), which yields $A=6.799 \pm 0.5637$ and $B=3.9669 \pm 0.4367$. To confirm the effectiveness of the two-spring model, the simulation results of $d_v/a_x$ are compared with the theoretical prediction of equation (\ref{equ:dv_dp}) in Fig. \ref{img:Result_of_double_spring_model}(a), which shows well consistence with an error less than $9\%$. This ensures the theoretical model will obtain diffusivity quantitatively agreed with the DPD simulation results.

As above mentioned, there are different opinions on the scaling law of the potential barrier with nanoparticle size in previous studies\cite{cai2015hopping,Dell2013Theory}.
Here, we try to clarify this issue by analyzing the contributions of the two kinds of deformation energy in our model.
In the double-spring model, the deformation energy of loop stretching $U_l \sim \chi \beta^{-1}$, and the strain energy induced by the junction deviation $W \sim \chi^{3} \beta$.
The ratio between the two parts of potential barrier $r_U = \frac{W}{U_l} \sim \chi^2 \beta^2$, which implies that $U_l$ will be dominated for smaller NP and more rigid network.
In this case, $U_l$ is linear varying with the nanoparticle radius, which is consistent with the prediction of Cai {\it et al.} \cite{cai2015hopping}, and inversely proportional to the stiffness of the polymer network. 
For larger $r_U$, the junctions deviation becomes more important than the loop stretching on the deformation energy, thus the potential barrier $U \approx W$ grows cubically with NP size or linearly with NP volume.
This is agreed with the result of Dell {\it et al.}\cite{Dell2013Theory}, i.e., linear increase of $U$ with NP size followed by a rapid increase. Moreover, contrary to its effect on $U_l$, the rigidity of the network will enhance the growth of $W$.

In summary, NP mobility in the unentangled polymer network with an ordered cubic topology is investigated by the coarse-grained molecular dynamics method for different NP size and network rigidity. The network rigidity is characterized by the normalized fluctuation length $d_{f}/a_x$ of the polymer strands. 
Results show that the junction deviation, which is measured by the averaged displacements of the junctions in the direction of the NP traversal, have profound contribution to the deformation of the polymer network, especially for larger nanoparticles and softer networks.
This kind of deformation is analyzed based on the phantom network model\cite{Flory1976Flory} in the conventional framework of continuum mechanics, which indicates that it can be simplified as a non-linear spring. 
Therefore, the double-spring model is presented to predict the effective diffusivity for the NP in polymer network, combining the mechanism of loop stretching proposed by Cai {\it et al.}\cite{cai2015hopping}, in which the related potential barrier is described as a linear spring.
The theoretical prediction has good agreement with our numerical results, and qualitatively consistent with previous theoretical study by Dell {\it et al.}\cite{Dell2013Theory}.
This model is found to be helpful to clarify the controversy over this issue in literature available\cite{cai2015hopping, Dell2013Theory}.
This theoretical framework is promising to be deepen our understanding on diffusion of nanoparticle in living systems with strong heterogeneity.

\bibliography{apssamp}

\end{document}